\RequirePackage{fix-cm}
\documentclass[twoside]{svjour3}       
\smartqed  
\usepackage{natbib}
\usepackage{array}
\usepackage{graphicx}
\usepackage{geometry}
\usepackage{amsmath}
\usepackage[fontsize=11pt]{scrextend} 
\usepackage[misc,geometry]{ifsym} 
\usepackage{authblk}

\usepackage{lineno}
\begin{document}

\title{Joint modeling of evacuation departure and travel times in hurricanes
}


\author{Hemant Gehlot$^1$   \and Arif M. Sadri$^2$ \and Satish V. Ukkusuri$^1$}


\institute{H. Gehlot (hgehlot@purdue.edu) \\ A. M. Sadri (asadri@fiu.edu) \\ S. V. Ukkusuri (\Letter) (sukkusur@purdue.edu) \at
              $^1$ Lyles School of Civil Engineering, Purdue University, West Lafayette, USA \\
              $^2$ Moss School of Construction, Infrastructure \& Sustainability, Florida International University, Miami, USA                      
}

\date{}

\maketitle

\begin{abstract}
Hurricanes are costly natural disasters periodically faced by households in coastal and to some extent, inland areas. A detailed understanding of evacuation behavior is fundamental to the development of efficient emergency plans. Once a household decides to evacuate, a key behavioral issue is the time at which individuals depart to reach their destination. An accurate estimation of evacuation departure time is useful to predict evacuation demand over time and develop effective evacuation strategies. In addition, the time it takes for evacuees to reach their preferred destinations is important. A holistic understanding of the factors that affect travel time is useful to emergency officials in controlling road traffic and helps in preventing adverse conditions like traffic jams. Past studies suggest that departure time and travel time can be related. Hence, an important question arises whether there is an interdependence between evacuation departure time and travel time? Does departing close to the landfall increases the possibility of traveling short distances? Are people more likely to depart early when destined to longer distances? In this study, we present a model to jointly estimate departure and travel times during hurricane evacuations. Empirical results underscore the importance of accommodating an inter-relationship among these dimensions of evacuation behavior. This paper also attempts to empirically investigate the influence of social ties of individuals on joint estimation of evacuation departure and travel times. Survey data from Hurricane Sandy is used for computing empirical results. Results indicate significant role of social networks in addition to other key factors on evacuation departure and travel times during hurricanes.
\end{abstract}

\section{Introduction}
\label{intro}
Hurricanes are among the mostly costly and dangerous natural disasters in the Unites States. In 2012, Hurricane Sandy caused 147 direct casualties along its path and brought damage in excess of \$50 billion for the United States \citep{nhc}. Other direct impacts of Superstorm Sandy include destruction of 570,000 buildings, cancellation of 20,000 airline flights, 8.6 million power outages in states among others \citep{weather}. These disastrous experiences have compelled various stakeholders such as emergency managers, crisis planners, researchers to uncover the critical role of evacuation logistics. However, effective evacuation planning is highly incumbent upon considering different interrelated steps of the overall evacuation process. A key behavioral issue is related to understanding and predicting evacuation departure time i.e. the time a household departs during an ongoing threat. 
Variations exist in the amount of time spent in deciding to evacuate, assembling family members and resources, and actually engaging in the act of leaving \citep{sorenson}. An accurate estimation of departure time will allow the prediction of dynamic evacuation demand and help develop effective evacuation strategies. Also, catastrophic consequences during a major hurricane can be significantly reduced by ensuring timely evacuations of vulnerable communities. Another important component of hurricane evacuations is the time it takes to reach preferred destinations and shelters. In this study, we refer to this time as evacuation travel time. During hurricane evacuations, the volume of vehicle traffic involved in mass evacuations consistently exceeds the capacities of local road networks, causing traffic jams \citep{dow}. By estimating the underlying factors that influence travel times, policy makers and evacuation officials can take suitable actions to reduce traffic jams. 

An important question that arises is that if the two components of evacuation are related to each other? Do people taking long journeys prefer to depart early in order to escape uncertain conditions like road jams? Does departing late leaves evacuees with less choices for destination and increases the probability to travel short journeys? Past studies in evacuation literature provide empirical evidences to these questions. It has been found that evacuation compliance is significantly higher among those who believe that they were given a mandatory evacuation order  \citep{baker1991hurricane,baker2000hurricane} and local evacuation notices are timed to allow all evacuating vehicles to clear the evacuation zone before the arrival of Tropical Storm wind. This indicates that late departure by evacuees (even though local officials issued a timely warning) may not leave enough time to travel distant places and evacuees would prefer to reach safe destinations near to their homes. In addition, evacuees who prefer to spend less time in traveling, also accordingly adjust their departure times based on their prior experiences \citep{noland2002travel,dalziell2001risk,rodrigo}. Previous studies like \cite{lindell2011logistics} and \cite{wu2012logistics} reported insignificant correlations between decision date and evacuation distance and \cite{wu2013logistics} found a significant negative correlation. The latter result suggests that the two variables can affect each other and this motivates us to explore joint modeling techniques. To the best of the authors' knowledge, this paper makes the first attempt to jointly model departure and travel times during hurricane evacuations. 

It has been documented in the past that social networks can serve as important contributors of disseminating hazard information. \cite{hasan} found that warning information propagates faster within networks with greater inter community connections. It has also been suggested that people first receive emergency warning information and then spread information to others \citep{sorenson,plg,perry1983citizen}. Sociological research has well documented that social networks serve the purpose of transmitting warning messages by disseminating information about an impending threat \citep{plg,sadri2017role}. For example, \cite{plg} documented the effect of peer contacts with warning content and
the number of warnings received.
Thus, an individual with more social connections can receive more warnings from his social connections. But at the same time empirical literature does not provide enough guidance about how social connections affect the two key behavioral issues being addressed: evacuation departure and travel times. Note that no one has explored the individual effect of social ties on these decisions, let alone the role of social ties on joint modeling. In this paper, we fill this gap by understanding the role of social ties along with other factors on joint estimation of departure and travel times. 

This study is organized in the following manner. Next section reviews the past studies related to evacuation departure and travel times and outlines the motivation for this study. Section 3 discusses the data source and sample used in empirical analysis. Section 4 advances the statistical framework for the joint model. Section 5 presents empirical results. Section 6 summarizes important findings from the research. The final section presents limitations of this study and provides some future directions.

\section{Background and motivation}
In the past few years, research in hurricane evacuation has increased substantially \citep{murray2013evacuation}. 
Researchers have addressed many decisions like evacuation decision \citep{dash,hasan2}, departure choices \citep{yin,vinayakj,lindell,sorenson,hasandep},  destination choice \citep{cheng, rodrigo}, route choice \citep{ivankat,sadriroute,sadriroute2}, mode choice \citep{sadrimode}, mobilization time \citep{sadrimob,vinayakmob}, and others. There are a modest number of studies on evacuation departure time but most of these focus on deriving empirical distributions without considering the influences of different factors. For a detailed review of evacuation departure timing studies, readers may refer to \cite{lindell}. On the contrary there have been few attempts in literature to develop behavioral models for evacuation timing decisions. For example, \cite{sorenson} made one of the earliest attempts using path analysis for evacuation timing behavior. Recently, \cite{hasandep} used a hazard-based modeling approach to model evacuation departure time. There are also a few studies that compute travel time during hurricane evacuations. In some works, travel time is estimated as an output of traffic flow models that use behavioral models like destination choice, route choice etc. \citep{lindell,pel}. Some works have also reported data on evacuation travel times and the additional time that it took compared to the normal time to travel to that destination \citep{wu2012logistics,wu2013logistics}. These studies reported the correlations of both variables with a variety of other variables.

Note that in hurricane scenarios, many evacuees depart by the time evacuation warning is issued \citep{lindell2005household,huang2012household,huang2016multistage}. It is possible that that people who depart close to the warning face less options in deciding their destinations and prefer safe destinations close to their home or alternatively some people decide to depart late because they have arranged to stay with peers or in commercial facilities close to home. It is also possible that limited options are likely to force late evacuees to travel farther because nearby accommodations have already been taken by earlier evacuees. Also, it has been observed that there is synchronization in the departure of people and many people evacuate during a particular period \citep{wolshon}. During this period, congestion on road networks increases. So, it is likely that evacuees would adjust their departure times based on prior evacuation experience to prevent delay in travel time. These arguments are also supported from past studies. It has been shown that the choice of departure time can be influenced by the time to reach the destination \citep{hasandep,deptime:diaz}. In addition, it is known that travel time depends on departure time due to dynamic nature of traffic flow and user demand \citep{merchant,han}. Thus, evacuation departure and travel time can be interrelated and affect each other. The final justification for considering an interrelationship comes from empirical results. This will be discussed in detail along with the empirical results later. 

An important aspect in the behavior of individuals during hurricane evacuation is the role of social ties. The role of social networks on hurricane evacuation decision behavior has been the focus of many works \citep{hasan,dash,famindis,sadriSN,sadriSN2}. \cite{plg} found that social networks (contacts with relatives and with friends and neighbors) had both direct and indirect effects on response to flood warnings. These studies suggest that individuals that have differences in their social network characteristics behave differently while deciding on whether to evacuate or not. But unanswered questions related to the role of social ties in joint modeling of evacuation departure and travel timings still remain. Note that the characteristics of social networks are important in individuals' decision making as the influence of social groups with different characteristics and demographics can significantly vary. Therefore, it is important to study the impact of aggregate characteristics of individuals' social networks on evacuation decision making. Do people with large social networks prefer to take long journeys? Does the age distribution of an individual's social network influences the individual's choice of departure time? We answer these questions in this paper. 

In summary, this paper makes the following contributions: 
\begin{itemize}
	\item Develops a model that considers interrelationship between evacuation departure and travel timings using joint modeling approach. 
	\item Establishes the significance of considering joint estimation of departure and travel times.
	\item Quantifies the role of social ties in joint estimation of evacuation departure and travel times. 
\end{itemize}


\section{Data source and sample selection}
The data set used in this paper was based on the responses of people affected during Hurricane Sandy. The sample frame for the survey was based on US census block group areas in coastal New Jersey and New York that have the likelihood of life-threatening storm surge in case of a major hurricane. Surveys were conducted on December 2-18, 2014 and January 5-15, 2015. The focus of the survey was on household evacuation decisions of people residing in the risk area. Tourists in the area on a short-term basis were thus not interviewed and were not included in the residence-based sampling frame. In addition, block groups located in New York City Evacuation Zone A not meeting the criterion of belonging to risk areas were also included. Approximately, 350 block groups were randomly selected from the list of qualifying block groups and a sample of telephone numbers was randomly drawn from those block group areas. Random sampling was done because the survey measured decisions about hurricane evacuation under risk, the sample had to enable inference to the population at risk for storm surge. Though the population resided in a relative small portion of the two states but inference still requires a random probability sample because of uncertainty associated with risk factor. In the sample of telephone numbers, 45\% were listed land line numbers and could be matched to block groups through address geo-coding. The remaining 55\% numbers were of cell phones that required additional pre-interview screening and verification through interview questions on zip code and distance to water body. Approximately 13\% of numbers dialed reached a household where an interviewer contacted an eligible person, and 40\% of those persons completed the survey. The average interview length was 15 minutes and the interviews were administered by trained bilingual interviewers (English and Spanish) who used questionnaires programmed in both languages on a computer-assisted telephone interview system. 

A total of 1100 households completed the survey. But out of these only 300 households evacuated during Hurricane Sandy. Note that only observations corresponding to households that evacuated are relevant to model departure and travel times. These 300 observations were cleaned of missing values and from observations where people did not respond, refused to answer or did not know about questions. After following this procedure, the cleaned data had 196 observations. The data set had variables relating to hurricane perception, evacuation and socio-economic characteristics of the respondents. Hurricane perception variables had questions to measure the threat perception of individuals. Evacuation variables capture the evacuation process of individuals by including the factors such as departure time, travel time, destination type etc. Socio-economic variables capture demographic and economic variables like gender, education, salary as well as social network variables like network size, age heterogeneity etc. 

As mentioned before, respondents were also asked about their social ties with the individuals with whom they interacted closely during Hurricane Sandy. These people constitute an individual's social network. Ego-centric social network data of individuals was obtained by following the personal network research design (PNRD) approach \citep{halgin}. The first step of PNRD is to create an exhaustive list of alters (social network members) with whom the ego (respondent) has some type of relationship. In the next step, respondents are asked about the attributes (for example, age and gender) of each alter, as well as the attributes of ego's relationship with alters (length of relationship, frequency of contact, and others). Based on respondents' replies, different measures of an individual's social network composition can be computed. One such measure is heterogeneity, which represents the variation in the values of an attribute across different alters. It will be discussed later in more detail. Table \ref{descstats} presents the descriptive statistics of explanatory variables (with abbreviations in italics) used in the final model specification with corresponding mean, standard deviation, minimum, and maximum. 

\begin{table}
	\caption{Descriptive statistics of explanatory variables}
	\begin{tabular}{>{}m{7.3cm}>{}m{1cm} >{}m{1.5cm}>{}m{1.7cm}>{}m{1.7cm}} \hline
		Variable and corresponding description& Mean& Standard deviation&Minimum&Maximum\tabularnewline \hline
		\textit{NJ}: Equal to 1 if household location is in New Jersey, 0 otherwise& 	0.58&	0.49&	0&	1\tabularnewline \hline
		\textit{Stormconcern}: Equal to 1 if an individual is concerned about the threat of a coastal storm, 0 otherwise 	&0.86&	0.35&	0&	1 \tabularnewline \hline
		\textit{Agehet}: Equal to 1 if age heterogeneity is 15 or over, 0 otherwise&	0.08&	0.27&	0&	1 \tabularnewline \hline
		\textit{Ordered\&sufinfo}: Equal to 1 if an individual perceives of being ordered to evacuate and had sufficient information to know what to do about Sandy, 0 otherwise &	0.44&	0.50&	0&	1 \tabularnewline \hline
		\textit{Old\&loctv}: Equal to 1 if older than 50 years and relied on local television stations for information, 0 otherwise&	0.55&	0.50&	0&	1 \tabularnewline \hline
		\textit{Household1\&reco}: Equal to 1 if household size is equal to 1 and the household perceives that it has been recommended to evacuate, 0 otherwise& 	0.05& 0.21& 0&	1 \tabularnewline \hline
		\textit{Loctv}: Equal to 1 if relied on local television stations for information, 0 otherwise &	0.88& 0.33& 0	&1 \tabularnewline \hline
		\textit{Np}: Size of social network (number of friends in an individual's social network)&	1.23&	1.76& 0&	6 \tabularnewline \hline
		\textit{Widow}: Equal to 1 if an individual is widow/widower, 0 otherwise&	0.12&	0.32&	0&	1 \tabularnewline \hline
		\textit{Sexhet}: Equal to 1 if sex heterogeneity (IQV) is 0.9 or over, 0 otherwise &	0.14& 0.35& 0&	1 \tabularnewline \hline
		\textit{Household1\&concern}: Equal to 1 if household size is 1 and the individual was concerned when Sandy was heading for the eastern coast of US, 0 otherwise& 	0.23& 0.42& 0&	1 \tabularnewline \hline
		\textit{Married\&evacbefore}: Equal to 1 if an individual is married and has evacuated before, 0 otherwise&	0.23& 0.42& 0&	1 \tabularnewline \hline
	\end{tabular}
	\label{descstats}
\end{table}

\section{Methodology}
\begin{figure}[h]
	\centering
	\includegraphics[scale=0.6]{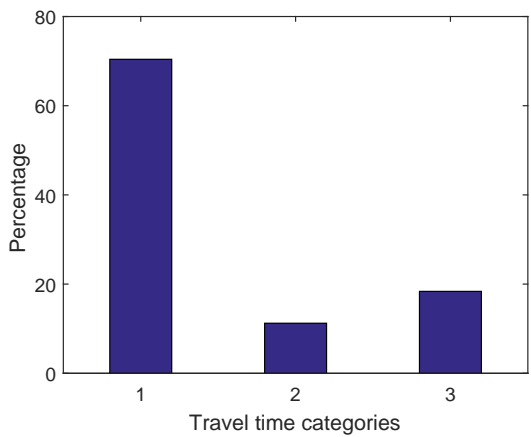} 
	\caption{Distribution of travel time (Category 1 is representing travel time being less than or equal to 1 hr, second category represents 1-3 hours and last category represents travel time of more than 3 hours)}
	\label{travtimdist}
\end{figure}

\begin{figure}[h]
	\centering
	\includegraphics[scale=0.33]{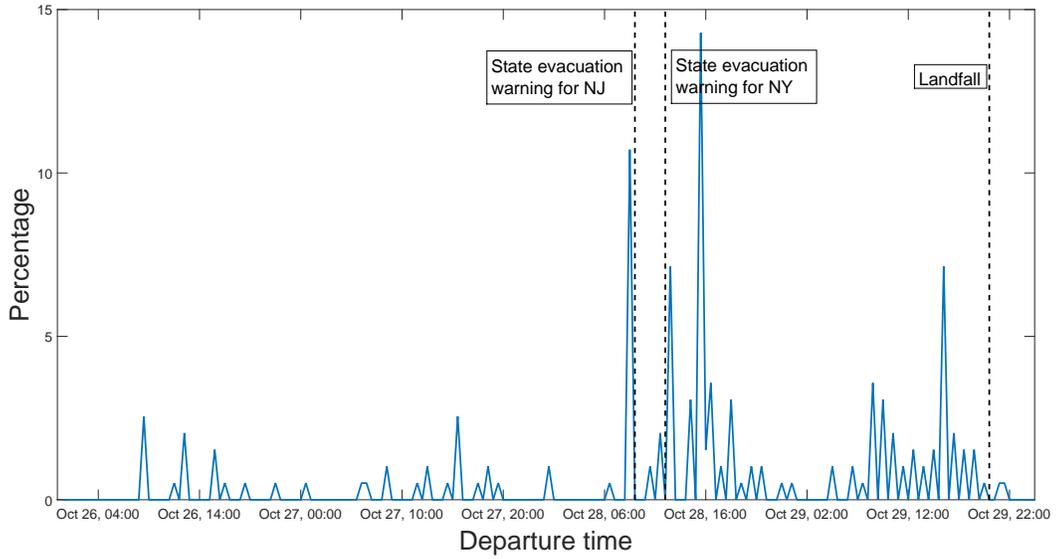} 
	\caption{Distribution of departure time (origin for departure time is approximately set as the time when the state of emergency was issued in different states). The state evacuation warnings of New Jersey (NJ) and New York (NY), and landfall of hurricane Sandy on the coast of NJ are represented by dashed lines.}
	\label{deptimedist}
\end{figure}

The methodology used here to model departure and travel times falls under the framework of discrete/continuous statistical models. We model departure and travel times as continuous and ordinal variables, respectively. The motivation to model travel time as an ordinal variable than as a continuous variable stems from the observed travel time distribution. Figures \ref{travtimdist} and \ref{deptimedist} present the distribution of travel and departure times, respectively, of evacuees in our final cleaned data. The first two categories represent intrastate travels and the last category represents interstate travels. We decided to have two categories for intrastate travels because a large section (more than 70\%) of the respondents reported their travel time as one hour and we wanted to have a more detailed truncation scheme. It was also observed that there are few non-one hour categories in the original data (non-categorized data). Hence, travel time data is concentrated to few discrete values. In addition, authors believe that truncation scheme should not make any significant difference since most of the observations had their travel times equal to one. Therefore, all the observations with travel time equal to one would always remain in the same category and at most, the difference due to truncation would only affect the travel time categories of 30\% observations. Next, same behavior is not observed for departure time distribution. The observed values of departure time are not concentrated (the highest percentage of any particular travel time value is 15\%) like those of travel time. Therefore, departure time is modeled as a continuous variable. 

There are two categories in discrete/continuous modeling framework: sequential and joint models. Sequential models use two stage estimation procedures on pooled samples \citep{train}. These models rely on the use of exogenous variables to replace variables that are thought to be endogenous and therefore correlated with the error terms of continuous equation. But a practical issue that emerges is the choice of exogenous variables, which may not be evident \citep{paez}. 

Joint discrete-continuous models make use of full information maximum likelihood (FIML) estimation techniques. In FIML, missing values are not replaced or imputed, but missing data is handled within the analysis model. \cite{bhat} developed a methodology to jointly model travelers' activity-type choice for participation, home-stay duration before participation in an out-of-home activity and out-of-home activity duration of participation. \cite{habibtoronto} explored the joint nature of mode choice and trip timing decisions using a joint discrete-continuous model. \cite{paez} proposed a ordinal-continuous model for analysis of activity episode generation and duration. 

In this study, we use a joint probability model that derives from individual discrete and continuous statistical models. A hazard-based model is employed to model departure time and travel time is modeled using ordered probit model. The structures and features of individual models and the joint model are discussed in the following three subsections.
\subsection{Continuous hazard duration model}
Continuous time hazard models are primarily concerned with the time or duration until the event time termination. In this paper, this is the time until one evacuates from the declaration of hurricane warning. There are two possible types of continuous time hazard specifications: proportional hazard model and accelerated time hazard model. The proportional specification targets the hazard rate, whereas the accelerated time specification targets the duration per se. An accelerated time hazard specification is the most appropriate approach in this case because it simplifies the likelihood function \citep{leetrb}. Assuming that the covariates act in exponential form, the accelerated time hazard model can be expressed in the following linear form for each observation:
\begin{linenomath*}
	\begin{equation*}
	\ln{d}=\gamma y+\alpha
	\end{equation*}
	\end{linenomath*}
where $d$ is departure time, $\gamma$ is a vector of estimable parameters, $y$ is a vector of explanatory variables and $\alpha$ is the unobserved error term. It is assumed that $\alpha$ is normally distributed with mean zero and standard deviation $\sigma$. The normal assumption on the distribution of error term makes our case a lognormal accelerated time hazard model. Note that normal assumption on error term simplifies the derivation of joint probabilities. Otherwise, error term should be converted into an equivalent standard normal variable so that correlation between the error terms of departure and travel times can be specified \citep{lee83}. So, probability density function, $f$, for departure time can be written as \citep{johnson}:
\begin{linenomath*}
	\begin{equation*}
	f(d)=\frac{1}{d\sigma}\phi \left( \frac{\ln{d}-\gamma y}{\sigma} \right)
	\end{equation*}
\end{linenomath*}
where $\phi (\cdot)$ denotes probability density function of standard normal distribution.
\subsection{Ordered probit model}
Ordered probability models are derived by defining an unobserved variable, $z$. This unobserved variable is typically specified as a linear function for each observation ($n$ subscripting omitted) \citep{washington}:
\begin{linenomath*}
	\begin{equation*}
	z=\beta x + \varepsilon
	\end{equation*}
	\end{linenomath*}
where $x$ is a vector of explanatory variables, $\beta$ is a vector of estimable parameters, and $\varepsilon$ is the unobserved error term. Using the previous equation, travel time (denoted by $tt$) for each observation is defined as:
\begin{linenomath*}
	\begin{equation*}
	tt=1 \hspace{34pt} \text{if} \hspace{6pt} z\leq \mu_0
	\end{equation*}
\end{linenomath*}
\begin{linenomath*}
	\begin{equation*}
	1 < tt \leq 3 \hspace{12pt} \text{if} \hspace{6pt} \mu_0 < z \leq \mu_1
	\end{equation*}
\end{linenomath*}
\begin{linenomath*}
	\begin{equation*}
	tt>3 \hspace{34pt} \text{if} \hspace{6pt} z > \mu_1
	\end{equation*}
	\end{linenomath*}
where $\mu_i$'s denote the threshold parameters that define $z$. In the above set of equations, $\mu_0$ is set to zero without any loss of generality. We assume the case of Probit model where error term is assumed to be normally distributed with mean zero and standard deviation one. The reason behind normal assumption is the same as that discussed for the error term of continuous hazard model. We define three categories $tt=1$, $1<tt\leq3$ and $tt>3$ as travel time categories 1, 2 and 3, respectively. So, the probability of travel time belonging to different ordinal categories can be written as \citep{washington}:
\begin{linenomath*}
	\begin{equation*}
	P(tt=1)= \Phi(-\beta x)
	\end{equation*}
	\end{linenomath*}
	\begin{linenomath*}
	\begin{equation*}
	P(1 < tt \leq 3)= \Phi(\mu_1-\beta x)-\Phi(-\beta x)
	\end{equation*}
	\end{linenomath*}
	\begin{linenomath*}
	\begin{equation*}
	P(tt>3)= \Phi(\beta x-\mu_1)
	\end{equation*}
	\end{linenomath*}
where $P(x)$ denotes the probability of event $x$ happening and $\Phi(\cdot)$ denotes cumulative distribution function of standard normal distribution.

\subsection{Joint model}
The joint estimation of accelerated time hazard model and ordered probit model requires error terms of both the models to be correlated. We define $\rho$ to be the correlation coefficient between the error terms. Let $d_1$, $d_2$ and $d_3$ denote the corresponding departure times for travel time categories 1, 2 and 3, respectively. Using conditional probability density identity, $f(A \cap B) = f(B) f(A|B)$, joint probability densities of departure and travel times belonging to different categories can be specified as:
\begin{linenomath*}
\begin{equation*}
f(d=d_1 \cap tt=1) = f(d=d_1)f(tt=1|d=d_1)
\end{equation*}
\end{linenomath*}
\begin{linenomath*}
\begin{equation*}
f(d=d_1 \cap 1 < tt \leq 3) =\\  f(d=d_2)f(tt \leq3|d=d_2) - f(d=d_2)f(tt<1|d=d_2) 
\end{equation*}
\end{linenomath*}
\begin{linenomath*}
\begin{equation*}
f(d=d_3 \cap tt>3) = f(d=d_3)f(tt>3|d=d_3)
\end{equation*}
\end{linenomath*}
As mentioned before, conditional probability densities in the above equations are evaluated by specifying correlation between the error terms of departure and travel times. To specify correlation between the two normal error terms we use conditional probability distribution of bivariate normal variables. If $X$ and $Y$ are normally distributed with mean $\mu_X$ and $\mu_Y$ and standard deviation $\sigma_X$ and $\sigma_Y$, respectively, then probability distribution of $Y$ given $X=x$ is \citep{bertsekas}:
\begin{linenomath}
\begin{equation} \label{conditional}
Y|(X=x) \sim N \left( \mu_Y+\rho \frac{\sigma_Y}{\sigma_X}(x-\mu_X) ,\sigma^2_Y \left(1-\rho^2\right) \right)
\end{equation}
\end{linenomath}
where $N(a,b)$ represents normal distribution with mean $a$ and variance $b$. Using conditional probability distribution in Equation \ref{conditional}, joint probability densities of different categories are as follows:
\begin{linenomath*}
\begin{equation*}
f(d=d_1 \cap tt=1) = \frac{1}{d_1\sigma}\phi \left( \frac{\ln{d_1}-\gamma y}{\sigma} \right)\Phi \left( \frac{-\beta x-\rho \left(\frac{\ln{d_1}-\gamma y}{\sigma}\right)}{\sqrt{1-\rho^2}} \right)
\end{equation*}
\end{linenomath*}
\begin{linenomath*}
\begin{multline*}
f(d=d_2 \cap 1<tt\leq3) = \frac{1}{d_2\sigma}\phi \left( \frac{\ln{d_2}-\gamma y}{\sigma} \right)\Phi \left( \frac{\mu_1-\beta x-\rho \left(\frac{\ln{d_2}-\gamma y}{\sigma}\right)}{\sqrt{1-\rho^2}} \right) - \\ \frac{1}{d_2\sigma}\phi \left( \frac{\ln{d_2}-\gamma y}{\sigma} \right)\Phi \left( \frac{-\beta x-\rho \left(\frac{\ln{d_2}-\gamma y}{\sigma}\right)}{\sqrt{1-\rho^2}} \right)
\end{multline*}
\end{linenomath*}
\begin{linenomath*}
\begin{equation*}
f(d=d_3 \cap tt>3) = \frac{1}{d_3\sigma}\phi \left( \frac{\ln{d_3}-\gamma y}{\sigma} \right)\Phi \left( \frac{\beta x-\mu_1-\rho \left(\frac{\ln{d_3}-\gamma y}{\sigma}\right)}{\sqrt{1-\rho^2}} \right)
\end{equation*}
\end{linenomath*}
Thus, joint likelihood function, $L$, is expressed as:
\begin{linenomath*}
	\begin{multline*}
	L=I(tt=1)\frac{1}{d_1\sigma}\phi \left( \frac{\ln{d_1}-\gamma y}{\sigma} \right)\Phi \left( \frac{-\beta x-\rho \left(\frac{\ln{d_1}-\gamma y}{\sigma}\right)}{\sqrt{1-\rho^2}} \right) + \\ I(1<tt\leq3)\frac{1}{d_2\sigma}\phi \left( \frac{\ln{d_2}-\gamma y}{\sigma} \right)\Phi \left( \frac{\mu_1-\beta x-\rho \left(\frac{\ln{d_2}-\gamma y}{\sigma}\right)}{\sqrt{1-\rho^2}} \right) - \\ I(1<tt\leq3)\frac{1}{d_2\sigma}\phi \left( \frac{\ln{d_2}-\gamma y}{\sigma} \right)\Phi \left( \frac{-\beta x-\rho \left(\frac{\ln{d_2}-\gamma y}{\sigma}\right)}{\sqrt{1-\rho^2}} \right) + \\ I(tt>3)\frac{1}{d_3\sigma}\phi \left( \frac{\ln{d_3}-\gamma y}{\sigma} \right)\Phi \left( \frac{\beta x-\mu_1-\rho \left(\frac{\ln{d_3}-\gamma y}{\sigma}\right)}{\sqrt{1-\rho^2}} \right)
	\end{multline*}
\end{linenomath*}
where $I$ is indicator function. For $N$ observations, log-likelihood function, $LL$, of the sample becomes
\begin{linenomath}
	\begin{equation}
	LL=\sum_{i=1}^N \ln{L}
	\end{equation}
\end{linenomath}
The log-likelihood function is non-convex in nature. Hence, it is optimized using the standard interior-point algorithm with BFGS hessian update in MATLAB \citep{coleman}. The standard errors of parameters are calculated using the inverse of Hessian procedure \citep{king}.

\section{Empirical analysis}
The choice of variables for potential inclusion in the model is guided by previous theoretical and empirical works on departure and travel times and intuitive arguments regarding the effect of exogenous variables on these variables. \cite{sorenson} had tested the influence of perceived threat on departure time of the people during the evacuation for a hazardous materials incident. Therefore, we include variable $Stormconcern$ to check its significance with the joint decision of departure and travel times during hurricanes. Note that hurricane Sandy made landfall in New Jersey so we decided to study the impact of being close to the landfall location on an individual's joint decision making by including the variable $NJ$. As mentioned in previous works like \cite{sadriSN2} social media sources like television have a significant impact on evacuation decision making therefore we decided to test its significance on departure time and travel time decisions by including the variable $Loctv$. People who are single or widowed also take into account considerations like safety while making evacuation trip decisions. Therefore, variable $Widow$ was checked for its significance in the joint model. In addition, as mentioned before social network parameters have been found to have significant effect on evacuation decisions \citep{sadriSN}. However, their effect on joint decisions of departure time and travel time has not been quantified in the past, therefore we include aggregated social network measures like $Np$, $Sexhet$ and $Agehet$ in our analysis. 
	
We also tested some interaction variables in our model. The variable $Ordered\&sufinfo$ is included because the evacuation behavior depends a lot on whether an individual believes if he/she has been ordered to evacuate or not \citep{huang2016leaves_SMA,baker1991hurricane}. Also, if an individual believes that he/she has been ordered to evacuate then does the knowledge of sufficient information have an impact on the evacuation patterns? Therefore, the effect of $Ordered\&sufinfo$ is also checked in addition to checking the significance of individual variables of this interaction variable. Note that many older people prefer using traditional sources of information like television more than the younger generation which prefers recent sources of information like social media \citep{kaplan2010users,grajczyk1998older}. Therefore, the inclusion of interaction variable $Old\&loctv$ offers new insights into the evacuation behavior of old people that prefer television. Also, single people have less constraints for evacuating as compared to people living with families. As we mentioned before, an order or recommendation plays an important role in evacuation patterns of an individual, we have an interaction variable $Household1\&reco$ to test the hypothesis such as do single people who are recommended to evacuate depart earlier than other people? Similarly, perceived risk can play a significant role in the evacuation patterns \citep{dash}. Therefore, we include the variable $Houeshold1\&concern$ to test hypotheses such as do single people who are more concerned towards hurricane risk prefer to travel short distances? Similarly, married households that have large number of people take into factors like amount of planning they need to do before planning their evacuation trip. Having the experience of evacuating before helps them to take an informed decision. Hence, we consider the interaction variable $Married\&evacbefore$ in our empirical analysis. 

Note that before testing the significance of various variables in the joint model it is ensured that endogenous variables and indicator variables having observations less than 15 are discarded. The former step prevents endogeneity and the latter step ensures that choices of the sample are representative of the population. We arrive at the final specification based on a systematic process of eliminating variables found to be statistically insignificant (with 90\% confidence level). Correlations between the explanatory variables are also analyzed and it is found that there is no pair of variables with high correlation. Thus, there is no issue of multicollinearity in the estimated model. The implications of the estimated results are presented in the subsequent sections. 
\subsection{Departure time model}
Table \ref{deptimres} presents the estimated results for departure time. First column presents variable name, second column provides the variable's estimated coefficient, third column provides the estimated coefficient's t-statistic and last column presents average marginal effect of the variable. Table \ref{deptimres_corelation} presents the correlation matrix of variables. It can be seen that no pair of variables has high correlation.   

Negative coefficient corresponding to \textit{NJ} implies that people from New Jersey departed earlier than the people from New York. It is likely that a household's departure time depends on its proximity from the projected path of hurricane. Since New Jersey was on the projected path of Sandy \citep{weather}, we observe early departure of New Jersey residents as compared to New York residents.   

Next, negative sign corresponding to \textit{Stormconcern} indicates that those who are concerned about the threat of coastal storms depart earlier than those who are not concerned. This finding is in accordance with a previous result that claimed that departure time is influenced by the seriousness of threat \citep{sorenson}. Also, negative coefficient corresponding to \textit{Ordered\&sufinfo} variable suggests that if people believe they are ordered to evacuate and have sufficient information on what to do when a hurricane strikes, then they depart earlier than the people who believe they are not ordered to evacuate or those who lack sufficient information. 
\begin{table}
	\caption{Departure time model estimates and average marginal effects}
	\begin{tabular}{>{}m{3cm}>{}m{2cm} >{}m{1.8cm}>{}m{3cm}} \hline
		Variable& Coefficient& t-stat&Average marginal effect\tabularnewline \hline
		Constant	&4.36&	38.60&	- \tabularnewline \hline
		\textit{NJ}&	-0.25&	-3.42&	-13.56 \tabularnewline \hline
		\textit{Stormconcern}&	-0.29&	-2.78&	-17.21 \tabularnewline \hline
		\textit{Ordered\&sufinfo} &  -0.18&	-2.43&	-9.58 \tabularnewline \hline
		\textit{Old\&loctv}&	0.16&	2.12&	8.47 \tabularnewline \hline
		\textit{Household1\&reco}&	-0.51&	-2.92&	-21.92 \tabularnewline \hline
		\textit{Agehet}&	0.22&	1.65&	12.77 \tabularnewline \hline
	\end{tabular}
	\label{deptimres}
\end{table}

Age of an evacuee and source of information can also have significant impact on departure time. Positive coefficient corresponding to \textit{Old\&loctv} indicates that people who are older than fifty years and rely on local television stations for information tend to depart late. It has been mentioned in previous works that old people are statistically more likely to have difficulty in moving quickly than young people \citep{bytheway}. There is also some evidence that mass media sources like television are quite effective in propagating warning information \citep{baker}. That is, a reliable source of information could prevent people from unnecessary early evacuation. Hence, the combined effect of age and information sources is a likely explanation behind the observed results. 
\begin{table}
	\caption{Correlation matrix for departure time model}
	\begin{tabular}{>{}m{2.3cm}>{}m{1cm} >{}m{1.8cm}>{}m{2.3cm}>{}m{1.3cm} >{}m{2.3cm}>{}m{0.8cm}>{}m{0.8cm}} \hline
		& \textit{NJ}&\textit{Stormconcern}&\textit{Ordered\&sufinfo}&\textit{Old\&loctv}&\textit{Household1\&reco}&\textit{Agehet}&\textit{DepTim}\tabularnewline \hline
		\textit{NJ}&	1&	-0.11&	-0.08&	0.08&	0.04&	-0.03&-0.22 \tabularnewline \hline
		\textit{Stormconcern}&	-0.11&	1&	0.13&	0.24&	0.02&	0.06&-0.17 \tabularnewline \hline
		\textit{Ordered\&sufinfo} &  -0.08&	0.13&	1&0.17&	-0.19&	-0.02&-0.12 \tabularnewline \hline
		\textit{Old\&loctv}&	0.08&	0.24&	0.17 &	1&	0.15&	0.03&0.01 \tabularnewline \hline
		\textit{Household1\&reco}&	0.04&	0.02&	-0.19 &	0.15&	1&	0.03 &-0.19\tabularnewline \hline
		\textit{Agehet}&	-0.03&	0.06&	-0.02 &	0.03&	0.03&	1&0.12\tabularnewline \hline
		\textit{DepTim}&	-0.22&	-0.17&	-0.12 &	0.01&	-0.19&	0.12&1\tabularnewline \hline
	\end{tabular}
	\label{deptimres_corelation}
\end{table}

Household characteristics like household size can also significantly impact departure timing. Negative coefficient corresponding to \textit{Household1\&reco} suggests that if there is only one person in a household and the household is recommended to leave then the household departs earlier than other households. The reasoning behind this observation is that people who live alone do not require time to organize other members of the household and can depart early. A recommendation to evacuate may increase the speed of deciding departure time and help in early departure. 

Social network characteristics also impact an individual's departure time. As mentioned before, heterogeneity is an important measure of an individual's social network. Higher the heterogeneity more the diversity among alters. We compute heterogeneity measures in E-Net, a network analysis program specifically developed for the analysis of ego-network data \citep{halgin}. For continuous variables like age, E-Net computes heterogeneity equal to the standard deviation of alters' values. Table 2 shows positive sign for the coefficient corresponding to the indicator variable for age heterogeneity. This implies that if standard deviation among alters’ age is 15 or more then ego departs late as compared to people whose age heterogeneity is less than 15. This might be because when age heterogeneity among alters increases beyond a threshold value there is delay in building the consensus for departure time which causes delay in ego's departure. This result can be used by policy makers to target groups of people with diverse age groups. If evacuation officials want early departure then they can specifically target these types of social groups. For instance, evacuation officials can prefer targeting residential areas where families having members from diverse age group live rather than targeting areas of less age heterogeneity like student accommodations.

The magnitude of the effect of explanatory variables on departure time can be examined by translating parameter estimates into marginal effects. Note that we report marginal effects in place of elasticities because elasticity measures are only defined for continuous variables and most of the estimated explanatory variables in our joint model are indicator values. Since marginal effects vary with observations, marginal effects are averaged over all the observations \citep{washington}. It can be seen that the magnitude of all the average marginal effects is less than 24 hours which implies that on average an explanatory variable can affect the decision to depart by utmost one day. The highest magnitude of marginal effect is shown by the variable corresponding to people who live alone and were recommended to leave. Negative marginal effect for this variable suggests that people who live alone can be targeted by evacuation officials. If evacuation officials recommend these people to leave, they would leave early and reduce the risk of facing hurricane. Instances of such type of people would be widows/widowers, divorcees/divorces etc. Marginal effect corresponding to those who were ordered to evacuate and had sufficient information can also be of interest to evacuation officials. If people have sufficient information and are ordered to leave then they depart earlier by 10 hours. Hence, policy makers should try to provide sufficient information to as many people as possible when they order them to evacuate.
\subsection{Travel time model}
This section discusses the estimated coefficients of exogenous variables for travel time. Table \ref{travtimres} presents the estimated results. First column presents variable name, second column provides variable's estimated coefficient, third column presents estimated coefficient's t-statistic and the last three columns present average marginal effects corresponding to different travel time categories. Table \ref{travtimres_corelation} presents the correlation matrix of variables. It can be seen that no pair of variables has high correlation. 

Positive coefficient corresponding to \textit{Loctv} suggests that those who rely on local television for information have more probability to travel for more than three hours than those who don't rely on local televisions. Here, local tv should be seen as a variable that influences travel parameters like destination. As mentioned before, local television stations provide reliable hurricane development information for local regions. If weather conditions are not favorable in that region then it is possible that evacuees would prefer traveling to distant places like evacuating to a different state. Thus, their probability to travel for more than three hours would increase.   

Individual characteristics like marital status can also significantly impact the travel time of an individual. Negative coefficient of indicator variable for widows/widowers suggests that these people are less likely to travel for more than three hours as compared to people of other marital status. That might be because widows/widowers don't feel safe and comfortable in traveling long distances. Hence, it is more likely for them to travel for one hour as compared to people of other marital status. Results also suggest that married people who have evacuated before for a coastal storm are less likely to travel for more than three hours as compared to other people. Since married people have larger households than other people they might not prefer to go on long journeys that require more planning and assembly of resources. If these people had evacuated before then they would be aware of safe locations near their house as compared to other people. Hence, they would prefer taking journeys of one hour than travel for more than three hours. Household characteristics, for example, size of a household also impacts travel time. Negative coefficient corresponding to \textit{Household1\&concern} suggests that people living alone who are concerned when a hurricane approaches have less likelihood of traveling for more than three hours. 

Social network variables like the size of social network and sex heterogeneity also indicate significant influence on travel time of an individual. Positive coefficient corresponding to \textit{Np} implies that as the size of an individual's social network increases the likelihood of traveling for more than three hours also increases. This might be because as the size of an individual's social network increases the probability to travel for more than three hours would also increase. This is because there might be consensus among the people in social groups to travel to the safest place among all the places suggested by the members of the group. This result can be used by emergency officials to target social groups that are small and do not travel enough distance to reach a safe location. Another significant social network variable is sex heterogeneity. Sex heterogeneity is represented using a measure called Index of Qualitative Variation (IQV) \citep{agresti}. IQV is a good measure of variability for nominal variables. IQV is based on the ratio of the total number of differences in the distribution to the maximum number of possible differences within the same distribution. For sex heterogeneity, it varies from 0 to 1 with 0 representing the lowest possible heterogeneity and 1 representing the highest possible heterogeneity. Negative coefficient corresponding to sex heterogeneity indicates that if an individual's social network has sex heterogeneity (IQV) of 0.9 or more then he/she has lesser likelihood of traveling for more than three hours than the people having sex heterogeneity less than 0.9. High heterogeneity might be representative of the presence of couples in an individual's social network. Note that couples might prefer to travel short distances for reasons like safety. Since, decisions of alters can also affect the decision of ego, an individual with sex heterogeneity higher than the stated threshold would have less likelihood of traveling for more than three hours. This result is helpful for emergency officials as they can make the shelter locations that are closer to people's homes as more suitable for accommodation of couples.  

Note that parameter estimates in an ordered probability model can only tell what will happen at the extreme categories (viz, travel time categories 1 and 3 in this study). Marginal effects provide more information on the magnitude of the change as well as what happens with interior categories (travel time category 2 in this study). For example, the value of average marginal effect of \textit{Loctv} for category 2 ($1<tt\leq3$) implies that if someone relies on local television then his/her probability to travel between 1 to 3 hours increases by 0.06. 

\begin{table}
	\caption{Travel time model estimates and average marginal effects}
	\begin{tabular}{>{}m{3.9cm}>{}m{1.8cm} >{}m{1.1cm}>{}m{2.0cm}>{}m{2cm}>{}m{2.0cm}} \hline
		Variable& Coefficient& t-stat&Average marginal effect (tt$=$1) &Average marginal effect (1$<$tt$\leq$3)&Average marginal effect (tt$>$3)\tabularnewline \hline
		Constant&	-0.95&	-2.86&	-&	-&	- \tabularnewline \hline
		\textit{Loctv}&	0.76&	2.27&	-0.20&	0.06&	0.14 \tabularnewline \hline
		\textit{Widow}&	-0.70&	-1.99&	0.19&	-0.05&	-0.14 \tabularnewline \hline
		\textit{Married\&evacbefore}&	-0.65&	-2.74&	0.19&	-0.05&	-0.14 \tabularnewline \hline
		\textit{Household1\&concern}&	-0.58&	-2.31&	0.17&	-0.04&	-0.13 \tabularnewline \hline
		\textit{Np}&	0.14&	2.29&	-0.10&	-0.02&	0.03 \tabularnewline \hline
		\textit{Sexhet}&	-0.64&	-1.92&	0.18&	-0.05&	-0.13 \tabularnewline \hline
	\end{tabular}
	\label{travtimres}
\end{table}

\begin{table}
	\caption{Correlation matrix for travel time model }
	\begin{tabular}{>{}m{2.8cm}>{}m{0.8cm} >{}m{0.8cm}>{}m{2.8cm}>{}m{2.8cm} >{}m{0.8cm}>{}m{0.8cm}>{}m{1cm}} \hline
		& \textit{Loctv}&\textit{Widow}&\textit{Married\&evacbefore}&\textit{Household1\&concern}&\textit{Np}&\textit{Sexhet}&\textit{TravTim}\tabularnewline \hline
		\textit{Loctv}&	1&	0.09&	0.01&	0.02&	-0.06&	-0.01&0.11 \tabularnewline \hline
		\textit{Widow}&	0.09&	1&	0.08&	0.29&	0.03&	-0.20&-0.14 \tabularnewline \hline
		\textit{Married\&evacbefore} &  0.01&	0.08&	1& 0.03&	0.30&	-0.12&-0.1 \tabularnewline \hline
		\textit{Household1\&concern}&	0.02&	0.29&	0.03 &1	&	0.03&	-0.25&-0.15 \tabularnewline \hline
		\textit{Np}&	-0.06&	0.03&	0.30 &	0.03&	1&	-0.07&0.1 \tabularnewline \hline
		\textit{Sexhet}&	-0.01&	-0.20&	-0.12 &	-0.25&	-0.07&	1&-0.12\tabularnewline \hline
		\textit{TravTim} &	0.11&	-0.14&	-0.1 &	-0.15&	0.1&-0.12&	1\tabularnewline \hline
	\end{tabular}
	\label{travtimres_corelation}
\end{table}

\subsection{Ancillary parameters and overall model fitness measures}
Table \ref{anc} presents the estimates of ancillary parameters, viz, standard deviation of hazard model, threshold parameter of ordered probit model and correlation coefficient. Out of these, most relevant parameter is the correlation coefficient, $\rho$, which is an estimate of correlation between the errors of departure time and travel time equations. That is because if this coefficient is statistically insignificant then there is no need to jointly model departure and travel times \citep{bhat}. The merit of joint estimation is that it captures unobserved correlations, which would be completely overlooked if independent models were used instead (causing the estimated parameters in independent models to be biased) \citep{habibtoronto}. Note that the results obtained from separately modeling the two decisions are not comparable to the joint model as significance of correlation coefficient confirms the significance of joint model. For instance, if we get significant correlation coefficient but still use separate models based on some performance measures like rho-square then we would be ignoring the correlation present between the unobserved factors of the two decisions.

Negative correlation indicates that unobserved factors that increase departure time of an individual also decrease the probability of travel time being more than three hours and increase the probability of travel time being equal to one hour of that individual and vice versa. For example, consider distance of household from a large water body as an unobserved factor that affects both departure and travel times. It is expected that if a household is close to a water-body, the members of the household will depart early to prevent the risk of facing dangerous conditions like floods. Also, household members would prefer traveling far enough from water-body to avoid any risk during hurricane. Another example is of people making decisions based on how long they have lived in their current county or city. It is expected that people who are relatively new to a place would be more uncertain about dangers related to hurricane and would depart early to take into account this uncertainty. Also, these people would consider traveling long distances because of unfamiliarity of safe locations in the vicinity. Thus, these examples are in agreement with the obtained result that says that if an unobserved factor decreases departure time then that factor increases the probability to travel for more than three hours. Note that one has to be careful when interpreting the values of correlation coefficients since they only capture the correlation between error terms that account for unobserved factors. In such cases, it is not possible to find out definite reasons behind the observed correlation value; we can only discuss likely reasons \citep{bhat,habibtoronto}. Nevertheless, the most important result is that there is significant correlation between departure and travel times and the proposed methodology incorporates this correlation. 

Table \ref{fitness} presents goodness of fit measures for the joint model system. A likelihood ratio test was conducted to test the goodness of the estimated model. The likelihood ratio test statistic is,
\begin{linenomath}
	\begin{equation}
	\chi^2 = -2[LL(r)-LL(\beta)]
	\end{equation}
\end{linenomath}
where $LL(\beta)$ is log-likelihood at convergence and $LL(r)$ is restricted log-likelihood. In restricted log-likelihood, exogenous variable parameters and correlation parameter are set to zero. It is found that $\chi^2$ is equal to 173.02 and is greater than the 99.99 \% confidence statistic, 26.22. This result clearly rejects the null hypothesis that exogenous variable parameters and correlation parameter are zero. 

Another measure of overall fit is adjusted $\rho^2$ statistic which is given as:
\begin{linenomath}
\begin{equation}
\text{Adjusted} \hspace{5pt}\rho^2= 1- \dfrac{LL(\beta)-K}{LL(r)}
\end{equation}
\end{linenomath}
where $K$ is the number of parameters for adjustment. It is equal to the difference in the number of parameters in $LL(\beta)$ and $LL(r)$. Adjusted $\rho^2$ for the estimated joint model is found to be equal to 0.1 which is considered reasonable for the current model given its complexity \citep{paez}. 
\begin{table}
	\caption{Estimated ancillary parameters}
	\begin{tabular}{>{}m{8cm}>{}m{2cm} >{}m{1.5cm}} \hline
		Variable& Coefficient& t-stat\tabularnewline \hline
		Standard deviation of accelerated hazard model $(\sigma)$&	0.49	&19.38 \tabularnewline \hline
		Threshold parameter of ordered probit model $(\mu)$	&0.41&	5.01 \tabularnewline \hline
		Correlation coefficient $(\rho)$&	-0.24&	-2.28 \tabularnewline \hline
	\end{tabular}
	\label{anc}
\end{table}
\begin{table}
	\caption{Summary of goodness of fit measures}
	\begin{tabular}{>{}m{7cm}>{}m{1.8cm} } \hline
		Measure&  Value\tabularnewline \hline
		Log-likelihood at convergence $(LL(c))$	&-1071.40 \tabularnewline \hline
		Restricted log-likelihood $(LL(r))$	&-1157.91 \tabularnewline \hline
		Number of parameters for adjustment $(K)$&	13 \tabularnewline \hline
		Number of observations&	196 \tabularnewline \hline
		Likelihood ratio test statistic $(\chi^2)$&	173.02 \tabularnewline \hline
		Adjusted $\rho^2$ 	&0.1 \tabularnewline \hline
	\end{tabular}
	\label{fitness}
\end{table}
\section{Summary and conclusions}
This paper presents a joint model for estimation of departure and travel times during hurricane evacuations. The methodology proposed in this paper is based on joint discrete-continuous modeling framework. To the best of authors' knowledge, this is the first study to jointly model departure and travel times during hurricane evacuations. A joint model accounts for potential interrelationship between two variables (in this study, evacuation travel and departure times) by incorporating correlation between the unobserved factors of these variables and estimating the log-likelihood function using full information maximum likelihood approach. The obtained empirical results emphasize the importance of jointly modeling these variables. We obtain a significant negative correlation coefficient for the joint model which implies that unobserved factors that increase the departure time of an evacuee also decrease the probability of an individual traveling for more than three hours. The model is based on a data obtained from a survey that interviewed households from New York and New Jersey who were affected by Hurricane Sandy. We also attempt to understand how social ties affect the combined evacuation timing preferences by collecting ego-centric social network information of survey respondents. 

Findings from this study contribute to a better understanding of departure and travel times during hurricane evacuations. Some of the key insights from this study include:
\begin{itemize}
	\item People living close to the projected path of hurricane evacuate earlier than those lying far away. This is an important finding because only a few studies have tried to relate departure time to location \citep{huang2012household}. \cite{hasandep} also found early departure time being associated with people who lived in the proximity to the projected path of the hurricane.  
	\item If individuals are concerned about the threat from coastal storms, then they depart earlier than the people who are not concerned. This observation is very helpful as departure time has not been related to threat perception in the past \citep{huang2012household}.
	\item Evacuees depart earlier if they believe they have been ordered to leave and have sufficient information about the hurricane than the people who don't belong to this category. That is, mandatory evacuation along with adequate information plays a positive role in early evacuation of people. \cite{lindell2005household} had observed significant relation between departure time and reliance on information from local authorities; and \cite{hasandep} observed early departure time being associated with people who received evacuation orders. Thus, we provide the combined effect of these two variables on departure time in our study.
	\item Individuals aging more than fifty years and relying on local television stations for information are likely to depart later as compared to those who don't belong to this category. The correlation of age and media sources like television on departure time has not been quantified in the past \citep{huang2012household}, thus, this study provides the effect of the combined variable on departure time. 
	\item If an individual lives alone and receives recommendation to evacuate, then his/her departure time decreases. Although the correlation of departure time with individual living alone has not been studied, recommendation to leave has been found to be correlated with early departure \citep{hasandep}.   
	\item If age heterogeneity among alters of an individual increases to 15 or more then the individual's departure time increases. In the past, effect of social network variables on departure time has not been studied \citep{huang2016leaves_SMA}. Thus, our study fills this gap.
	\item The probability to travel for more than three hours increases if someone relies on local television for information during hurricane period. In the past, the relation of using social media sources on travel time has not been documented and this paper fulfills that gap. 
	\item Widows/widowers are less likely to travel more than three hours as compared to individuals with other marital status. Also, married people who have evacuated before for a coastal storm are less likely to travel for more than three hours as compared to the rest of people. \cite{wu2012logistics} had reported that married evacuees travel shorter travel durations. Thus, our study confirms with this finding and provides the relation of the combined variable on travel time.
	\item Those who live alone and are concerned when a hurricane approaches have less likelihood of traveling for more than three hours than other people. The correlation of this combined variable has not been studied before, so our study is an important step in that direction.
	\item As the size of an individual's social network increases the probability to travel long trips (more than three hours) increases. If sex heterogeneity (IQV) among alters is 0.9 or above then probability to travel for more than three hours decreases. The relation of social network variables on travel time has not been studied before and this study fills the gap in this area.
\end{itemize}
The above insights provide important implications for effective evacuation planning. For example, evacuation planners can target specific groups like those who live alone and recommend them to leave. These people would then depart earlier than had they not been recommended. In this way, evacuation planners can control the movement of traffic and control road jams. 

\section{Limitations and future directions}
There are some limitations of this study that should be of interest while extending the results to other scenarios and for conducting future research. One limitation of the study is that it is based on Hurricane Sandy for which the state governments had issued local evacuation orders although federal evacuation warning from National hurricane center was not issued. In some cases, such as Hurricane Harvey, there were no evacuation orders for millions of Houston-area residents \citep{harvey}. Another limitation is that we do not compare our results obtained with other data sources like Twitter. The comparison of results from traditional sources like survey with recent social media platforms like Twitter would provide more insights into evacuation behaviors. In addition, this study did not take into account the behavior of tourist population residing at the time of hurricane and can be done in future in the lines of general disaster setting works \citep{tourist}. Also, due to the complexity of the model we assumed a fixed parameter model, however, random parameters can be introduced that would account for the heterogeneity in the estimated parameters across different observations. The joint nature of other evacuation decisions like departure time-route choice, departure time-destination choice can also be explored in the future. In addition, if a similar data for a different hazard is collected in future then transferability of the results should be checked. \\

\small
\noindent \textbf{Acknowledgements}
The authors are grateful to National Science Foundation for the grant CMMI-1131503 to support the research presented in this paper. A number of questions used for the survey questionnaire were derived from earlier research on the Hurricane Sandy evacuation done by Dr. Hugh Gladwin and Dr. Betty Morrow supported by National Science Foundation grants CMMI-1322088 and CMMI-1520338. The survey was conducted by Dr. Hugh Gladwin of Florida International University. However, the authors are solely responsible for the findings presented in this study. \\

\noindent \textbf{Author Contributions Statement} All the authors have contributed to the design of the study, conduct of the research, and writing the manuscript. All authors gave final approval for publication. \\

\noindent \textbf{Competing Financial Interests} Authors declare no competing financial interests.

\normalsize
\bibliographystyle{spbasic}      
\bibliography{LaTeX}   

\end{document}